PERSPECTIVE

# Self-Organization Dynamics Beyond Equilibrium: Discreteness, Computation, and Rules of Life

Hyun Youk[1,2,3]

[1]Department of Systems Biology, University of Massachusetts Chan Medical School, Worcester MA, USA
[2]Department of Physics, University of Illinois Urbana-Champaign, Urbana IL, USA
[3]Correspondence: youk@illinois.edu

**SUMMARY**

**Living systems self-organize in ways that conventional physical frameworks—based on forces, energies, and continuous fields—cannot fully capture. Processes like gene regulation and cellular decision-making involve rule-based logic and computational interactions. Here, I introduce the concept of non-equilibrium capacity (NEC) to denote the finite capacity of living systems to generate and sustain life-associated dynamics—the very capacity that defines viability—and whose irreversible loss constitutes death. I argue that two lines of inquiry are especially promising for understanding why this capacity is inevitably lost. First, experiments that slow or suspend all cellular processes reveal "low speed limits" below which life collapses. Second, generalized cellular automata—where cells interact over diffusion-defined neighborhoods and obey discrete rules—provide a framework to understand how order emerges or persists. Together, these approaches suggest a new grammar of biology that complements energy-based physics and explains how living systems sustain and ultimately lose their NEC.**

## What is self-organization?

A system self-organizes when it spontaneously evolves from a disordered configuration into an ordered one[1-3]. This deceptively simple statement raises deep questions: how should we quantify order? What determines the timescales and pathways by which disorder gives way to order? Why are some self-organized states static while others remain dynamic and continually renewed? And, most profoundly for biology, why do such dynamics persist only for finite durations before ultimately decaying to thermal equilibrium[4]—that is, to death? Although self-organization occurs in both non-living and living systems, these issues take on a different urgency in biology: for living systems, understanding how order emerges, is sustained, and ultimately fails is central to understanding life itself.

In this Perspective, I focus on why self-organization in living systems poses challenges that conventional physical frameworks struggle to meet. Force- and energy-based descriptions excel at capturing motility, mechanics, and interactions in non-living matter, but many biological rules are qualitatively different: genes switch on or off, proteins bind or unbind, cells secrete or sense. These rules are discrete, computational in character, and not naturally expressible in terms of forces or fluxes. At the same time, living systems confront us with questions that equilibrium or near-equilibrium physics does not. Why is the capacity for sustaining non-equilibrium dynamics always finite? Why is viability inevitably and irreversibly lost? How do discrete local rules scale into collective organization? Addressing these challenges may require new metrics and concepts—abstractions that, like entropy or temperature in physics, condense vast microscopic details into variables that reveal universal principles. Developing such a grammar would not only deepen our understanding of life but also broaden the scope of physics itself.

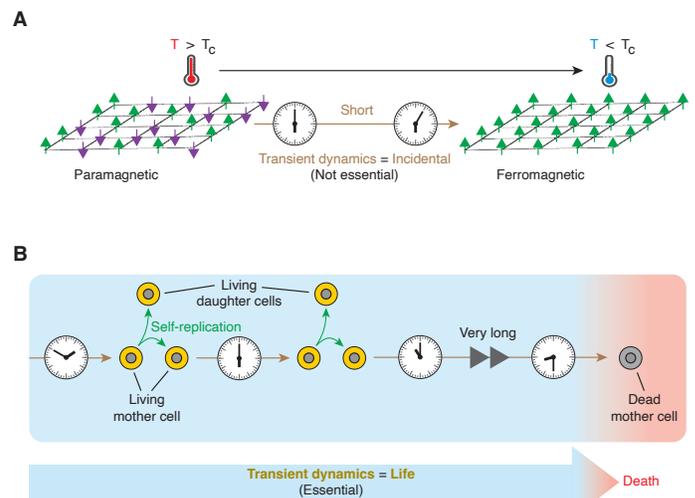

**Figure 1. Transient dynamics in non-living versus living systems.**
*(A)* A phase transition in a two-dimensional Ising system occurs when temperature is quenched below the critical value Tc. The system evolves from a paramagnetic to a ferromagnetic state through a short transient, which is incidental and not essential for understanding the two equilibrium states. *(B)* In contrast, a living cell (yellow circle with grey nucleus) continuously exhibits non-equilibrium dynamics—growth, replication, and information processing—until it dies and reaches thermal equilibrium. Here the transient dynamics are essential: they constitute the state of being alive itself.

## Limits of Equilibrium Phase-Transition Frameworks

The modern study of self-organization originates in the physics of non-living matter. Superfluidity in helium and ferromagnetism in paramagnets are iconic examples[5-7]: order emerges when an external parameter, such as temperature, is tuned across a critical point (Fig. 1A). Condensed-matter physics has developed a powerful toolkit here—order parameters, quasi-particles, symmetry classifications, and singularities in thermodynamic functions[8,9]. These concepts explain phase transitions across diverse materials, testifying to the remarkable generality of equilibrium statistical mechanics.

Yet these frameworks rarely capture the real-time dynamics of phase transitions. For example, quenching a classical spin system across a critical point can lead to domain growth and defect annihilation, phenomena typically studied through surrogate, phenomenological dynamics obtained with the Metropolis updates or Langevin dynamics rather than exact equations of motion[3]. In quantum systems, a sudden quench drives a many-body Hamiltonian into non-equilibrium evolution that, in principle, requires solving the time-dependent Schrödinger equation for a Hilbert space whose dimension grows exponentially with system size. While cold-atom experiments provide glimpses of the spatiotemporal evolution[10-14], no general theory exists. Beyond broad scaling arguments, as in the Kibble–Zurek mechanism, obtaining spatiotemporal pathways from a disordered equilibrium phase to an ordered one—and vice versa—remains an open challenge[15-19].

For many non-living systems, one could argue that knowing the real-time self-organization dynamics is unnecessary: the transient trajectory is incidental, because the system inevitably relaxes to equilibrium, and in non-glassy materials this happens on relatively short timescales (Fig. 1A). In living systems, by contrast, thermal equilibrium is synonymous with death—even though cells can die long before reaching it. For living systems, the transient dynamics are not incidental but fundamental: they represent the state of being alive and are the very processes we seek to understand (Fig. 1B).

## Limitations of Dissipative-Structure Framework

Foundational advances in non-equilibrium physics are essential for understanding the self-organization dynamics that mark and sustain living systems. At the same time, living systems themselves may be a testbed for those advances. Early progress came in the 1930s with Lars Onsager, whose reciprocal relations explained how small, linear deviations from thermal equilibrium can sustain steady fluxes—of heat, particles, or charge—and thereby yield non-equilibrium steady states[20,21]. But Onsager's framework applies only near thermal equilibrium, where a system responds linearly. In the 1960s and 1970s, Ilya Prigogine and Grégoire Nicolis pushed further. They showed that far from thermal equilibrium, nonlinear feedbacks under continuous fluxes of energy and matter can produce stable, spatially organized "dissipative structures"[22,23]. Examples include Bénard convection rolls, Belousov–Zhabotinsky oscillations, and reaction–diffusion waves, which persist indefinitely while sustained by fluxes of matter and/or energy (Fig. 2A). Crucially, they identified two ingredients necessary for the self-organization of dissipative structures from an initially homogeneous state—whether a uniformly mixed chemical pool or a fluid of uniform density: nonlinear dynamics and sustained external constraints that keep the system far from thermal equilibrium (Fig. 2B).

Prigogine and Nicolis's landmark generalization was to explain how order could be a steady state far from thermal equilibrium. Yet for living systems, their framework falls short. Dissipative structures are described by smooth fields of matter density and chemical concentration, whereas cells, genes, and molecules are discrete entities, often few in copy number, whose individuality matters and cannot be reduced to mathematically continuous functions. Moreover, dissipative structures exist indefinitely—they are perpetually recreated—when externally driven, but living systems do not: all cells and organisms die, even with ample nutrients and habitable conditions (Fig. 2C). Even seemingly cyclical processes such as the cell cycle are never exact repeats at molecular resolution, since not every molecule is in the same position or copy number as in the previous cycle, owing to stochastic fluctuations. Most importantly, unlike the non-living dissipative structures explained by Prigogine and Nicolis, living cells and organisms adapt, compute, process information, and replicate. Their dynamics are actively reshaped by information, not merely by passive fluxes of matter and energy. Thus, while the dissipative-structure framework greatly advanced non-equilibrium physics, it has not captured the transient, agent-based, information-responsive dynamics that define life—a mismatch rooted in the very grammar of physics, whose metrics and concepts remain poorly fitted to biological agents and rules.

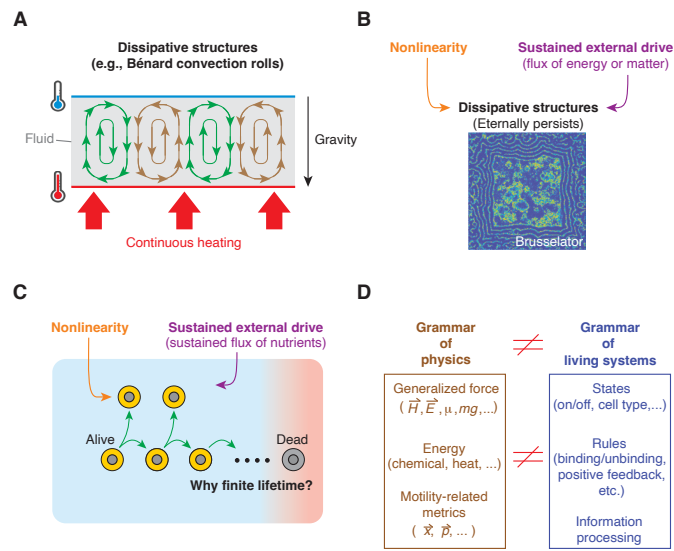

**Figure 2. Dissipative structures versus living systems.**
*(A)* Classical dissipative structures such as Bénard convection rolls arise when a fluid heated from below becomes unstable, breaking symmetry and forming persistent, organized patterns. *(B)* Two ingredients underlie such structures: nonlinear dynamics and sustained external fluxes of energy or matter. As long as these fluxes continue, dissipative structures can persist indefinitely. *(C)* By contrast, living systems also rely on nonlinearity and sustained fluxes (nutrients, energy), yet their dynamics have a finite lifetime: cells eventually die, even under favorable conditions. *(D)* This limitation reflects a deeper mismatch between the grammar of physics and that of biology. Physics describes systems in terms of generalized forces, energies, and motility-related variables, whereas living systems operate through discrete states, rules, and information-processing events such as gene regulation and signaling.

In recent decades, stochastic thermodynamics has extended Prigogine's program by applying thermodynamic reasoning to small, fluctuating systems such as single molecules, enzymes, or synthetic nanodevices[24]. It formulates fluctuating trajectories of microstate quantities like entropy production or free-energy dissipation, and it has revealed universal constraints such as fluctuation theorems and thermodynamic uncertainty relations[25,26]. Yet despite these successes, stochastic thermodynamics retains the grammar of classical thermodynamics: it is still written in terms of energies, forces, and fluxes. This grammar, powerful for colloids or biomolecular motors, remains poorly suited for describing living cells and organisms, where rules are expressed in terms of gene expression regulation, cellular decisions, and information processing (Fig. 2D). Even in its modern stochastic form, thermodynamics cannot fully capture the self-organization dynamics that define life.

## Challenges posed by self-organization dynamics that sustain and define life

What is missing in our current understanding of self-organization dynamics that define life? I propose that it is not enough to distinguish non-equilibrium systems from equilibrium ones. Instead, we must ask why the self-organization dynamics of a cell—the simplest unit of life—cannot continue indefinitely, even with endless nutrients and a permanently habitable environment. Being out of thermal equilibrium is necessary for life, but it is not sufficient: a dead, decomposing cell is also out of thermal equilibrium.

What distinguishes living systems is their capacity to generate and sustain non-equilibrium dynamics associated with life, such as self-replication, self-organization, and information processing. This non-equilibrium capacity (NEC) persists even in extreme states where cells appear lifeless, such as when frozen, desiccated into powder, or existing as dormant spores (Fig. 3)[27-31]. While viability in microbiology has long been—and remains—defined by a cell's capacity to self-replicate—captured, for instance, by colony-forming assays—many living entities, including non-dividing or multicellular systems, are alive and responsive even if they have lost the capacity to replicate sexually or asexually. For this reason, I use NEC as a unifying concept that encompasses both replicative and non-replicative living systems: the ability to generate, sustain, and renew non-equilibrium dynamics such as gene expression, information processing, and self-organization.

The NEC also extends our focus beyond dynamics currently underway—replication in nutrient-rich conditions, metabolism in progress—to the latent potential to restart them. Yet it is unclear how the static configuration of molecules inside a cell at one moment determines whether that cell retains the capacity to resume and sustain the processes that make it alive. I propose that a fundamental question—one that has received relatively little attention—is why and how the NEC that defines living systems is irreversibly and inevitably lost.

A promising first step to answering this question is to determine the limits of self-organization dynamics that sustain viability by deliberately slowing or suspending all cellular processes, as a few recent studies have done. In one study, lowering temperature to near 0 °C slowed all molecular processes in yeast—a model eukaryotic cell—without completely freezing it. This revealed that there is a slowest pace at which life can progress, and that attempting to go slower causes the cell to rupture and thereby irreversibly lose its NEC[32]. In other words, for each temperature there exists a "low speed limit" below which a yeast cell cannot remain viable: unlike a dormant spore that can pause replication yet stay viable (Fig. 3), here the cell's attempt to indefinitely delay replication leads to physical collapse. Another study slowed the replication dynamics of embryonic stem cells by differentiating them into specialized cell types, and likewise showed that growth cannot be reduced to an arbitrarily slow pace without killing the cells[33]. Both studies point to a broader conceptual framework—one echoed in other studies of cell death[34,35]—in which two opposing processes compete to determine viability: order-promoting processes such as gene expression that build or repair the cell, and order-demoting processes such as creation of reactive oxygen species[32]—inevitable byproducts of metabolism that damage cells. Death occurs when the balance tips irreversibly toward disorder.

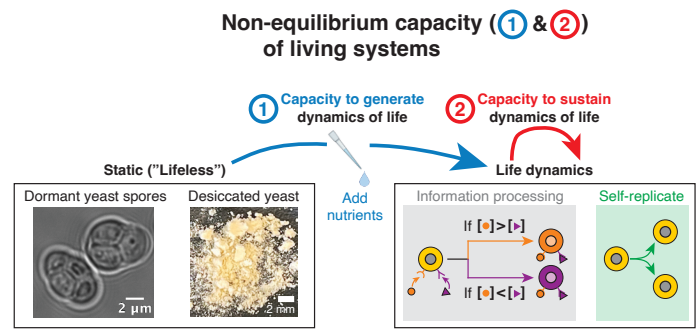

**Figure 3. Non-equilibrium capacity (NEC) of living systems.**
*Even in extreme conditions where cells appear lifeless—frozen, desiccated, or dormant in water—they may retain a latent capacity to restart replication and metabolism. This non-equilibrium capacity (NEC) distinguishes living systems from dead matter: it extends beyond ongoing processes to the potential to resume them in the future. Unlike dissipative structures in non-living matter, however, NEC is finite and ultimately irreversibly lost, even under favorable conditions.*

These slowing and suspending approaches, while still in their early stages, offer a way to probe not only the boundaries of viability but also the deeper question of why death is irreversible and inevitable. By pushing cells toward their "low speed limits," we can make explicit the competition between order-promoting and order-demoting processes. So long as the effective rate of order-promoting processes exceeds that of order-demoting ones, the cell remains viable. But because order-demoting pro-

cesses—such as damage from metabolic byproducts—may not be fully eliminated, only counteracted, and may leave permanent damage that cannot be completely reset, eventually there may come a point when their effective rate surpasses that of the order-promoting processes. This crossover may also arise from stochastic fluctuations in the rates and from the inherent heterogeneity of cells: no cell ever perfectly retraces the molecular configurations that once allowed order-promoting processes to dominate. Through these multiple routes, the system reaches a tipping point at which the capacity for non-equilibrium dynamics is lost irreversibly, regardless of the specific molecular details. Thus, slowing life does more than reveal boundary conditions: it provides an experimental foothold for understanding why the loss of viability is both irreversible and inevitable.

## Beyond the Deluge of Parameters: Toward New Metrics

To make further progress on understanding why viability-sustaining dynamics cannot indefinitely persist, we must confront a more foundational mismatch: the grammar of physics—its reliance on forces, energies, and motion-related variables—fits poorly with the logic-like, algorithmic control of biological regulation (Fig. 2D). These physical quantities do not capture processes such as gene regulation, where the specificity of DNA sequences, protein structures, and chemical reactions is central. When a relevant quantity cannot even be meaningfully defined, a practical strategy is to reformulate the problem in terms of other, measurable quantities so that predictions can be experimentally tested. From a more philosophical viewpoint, this may require abandoning entrenched concepts and inventing new ones, much as quantum mechanics forced a reconceptualization of physical state, position, and momentum. Similarly, although physical notions such as energy and momentum surely exist in cells, they are not the right variables for describing phenomena like gene expression or cell-state transitions. Alternative quantities—such as information in the Shannon sense—offer partial insight by assessing how close a system comes to optimal performance, and living systems have indeed motivated novel applications of information theory. Yet information measures primarily assess how well signals are transmitted under constraints; they do not by themselves specify how discrete, rule-executing dynamics construct and sustain ordered states over time. Information measures bound performance; they do not construct dynamics. Hence, even with this framework, we remain unable to predict or explain many dynamics of cells, genes, and their collectives. For example, we still lack a general language or set of concepts for describing the dynamics of spatial self-organization by hundreds of communicating cells.

The prevailing strategy has instead been to describe each chemical or cellular process—such as the expression of a gene—with a differential equation, parameterized by rate constants. This framework has achieved notable successes: it explained E. coli chemotaxis[36], cAMP signaling in Dictyostelium discoideum[37], and biofilm dynamics[38], and by extension inspired Whole-Cell Models (WCMs) that simulate every molecule in a bacterium or synthetic minimal cell[39,40]. By discretizing space into voxels and running stochastic reaction networks inside each voxel, WCMs integrate diverse datasets into simulations that reproduce global, self-organized properties such as cell growth and division times. These are technical triumphs and powerful computational experiments—especially valuable because we cannot simultaneously measure everything inside a real cell. At the same time, WCMs underscore a gap: simulation alone does not automatically yield higher-level variables or generalizable principles. Just as tracking every ideal-gas molecule does not by itself produce the concept of temperature, simulating every intracellular process does not by itself generate the kinds of abstractions that define life. Rate-based models also remain tailored to the specific systems from which their parameters are measured and therefore do not readily generalize across organisms or contexts.

Rather than viewing these approaches as competing, I suggest they are complementary. WCMs offer a "maximalist" strategy—reconstructing molecular detail to test integrative hypotheses—while concept-seeking approaches aim for a "minimalist" strategy—abstracting away detail to reveal governing rules. Both are essential, and progress in understanding living systems will require them to inform each other.

## Discreteness and Cellular Automata as a Framework

One natural route toward such minimalist, concept-seeking approaches is to embrace the discreteness that is intrinsic to living systems. Discreteness arises at multiple levels: agents (cells are countable, not continuous densities), states (a gene is expressed or not; a protein binds or unbinds), space (cells occupy specific positions rather than being smeared over continua), and time (cellular events occur in steps, often separated by orders of magnitude in timescales). For decades, biologists have described systems in these terms—qualitatively, through logic-gate-like rules such as "this gene switches on when that protein is present." Such descriptions are discrete at their core and, I propose, reflect the natural grammar of biology—and its computational character—more faithfully than continuous equations of motion with real-valued rate constants. This view echoes recent calls to incorporate ideas from complexity and information theory into biology[41].

Cellular automata (CA) provide a unifying computational framework for this discreteness. In their general form, CA treat agents, states, space, and time as discrete, and update local rules synchronously or asynchronously to produce global dynamics—allowing us to watch self-organization unfold, not merely infer final patterns. Historically, related models mapped out special regions of this design space. Alan Turing's 1952 the-ory formulated continuous reaction–diffusion equations and used linear stability analysis to show how periodic patterns can emerge from a homogeneous state[42]; later studies explored lattice discretization for analysis and simulation[43]. Coupled-map lattices discretized space and time while keeping states continuous, illuminating spatiotemporal chaos and synchronization[44]. Lattice-gas cellular automata (LGCA) represent each lattice node with channels corresponding to discrete velocity directions; particles occupy channels and update through alternating propagation and collision steps[45] (Fig. 4A). From these microscopic rules, mean-field approximations yield lattice Boltzmann equations, which in turn recover hydrodynamic equations in the continuum limit. Variants of LGCA have been adapted to model microbial chemotaxis and collective motility. More biologically focused models, such as the Glazier–Graner Cellular Potts Model[46], represent tissues as collections of discrete cells whose rearrangements reflect differential adhesion and other mechanical interactions. Each of these approaches captures aspects of discreteness—especially motility and mechanics—but largely retains the physics grammar of forces, energies, and motions.

By contrast, CA beyond motility let us encode biological rules in their native, qualitative form: genes switching on or off, cells secreting or sensing molecules, and local updates triggered by discrete conditions. This makes CA especially suited for modeling how intracellular and intercellular rules coalesce into emergent self-organization dynamics—from static spatial patterns to persistent wave-like activity. Complementary to exhaustive simulations such as WCMs, which aim to reconstruct molecular detail, CA offer a minimalist, rule-based grammar that can reveal principles without requiring every parameter. Historically, this is precisely the ambition that led to the invention of CA: John von Neumann, inspired by Stanislaw Ulam, designed cellular automata to uncover minimal rules for self-replication, and John Conway's Game of Life showed how deceptively simple update rules can generate extraordinary computational richness. Early theoretical efforts by von Neumann[47] and Langton[48] demonstrated that even simple discrete update rules can sustain or fail to sustain self-replicating structures—offering conceptual parallels to how living systems maintain or lose their NEC. These foundations set the stage for modern extensions of CA, which aim to incorporate biological communication and multicellular organization more explicitly. These models, by making explicit when discrete, rule-based dynamics sustain or fail to sustain organized behavior, offer a way to probe the boundary between viable and non-viable dynamics—the computational analogue of how living systems either preserve or lose their NEC.

## Generalized Cellular Automata for Multicellular Self-Organization Dynamics

Researchers have recently adapted cellular automata to model the grammar of cells, particularly how they communicate through non-mechanical means—for example, by secreting and sensing diffusible molecules[49,50]. To capture such biological communication more faithfully, these models extend the classical nearest-neighbor setting. These extensions, which we can call "generalized CA", allow cells to interact not only with immediate neighbors but also across diffusion-defined neighborhoods, as occurs in real tissues. An impetus for this rejuvenation of CA is the accumulation of knowledge from synthetic and systems biology, which has uncovered wiring diagrams of gene networks and quantified the strengths of regulatory interactions. Just as von Neumann was inspired by biological self-replication in designing his original automata, but unlike his abstract rules, generalized CA can now be built directly from experimentally measured interactions. This shift enables two goals: (1) by reducing experimentally derived communication rules to their bare minimum, determine which minimal interactions suffice to produce observed dynamics; and (2) explore the computational character of biological rules themselves—potentially opening new avenues for understanding the physical basis of computation, much as biological principles have previously inspired algorithms in computer science.

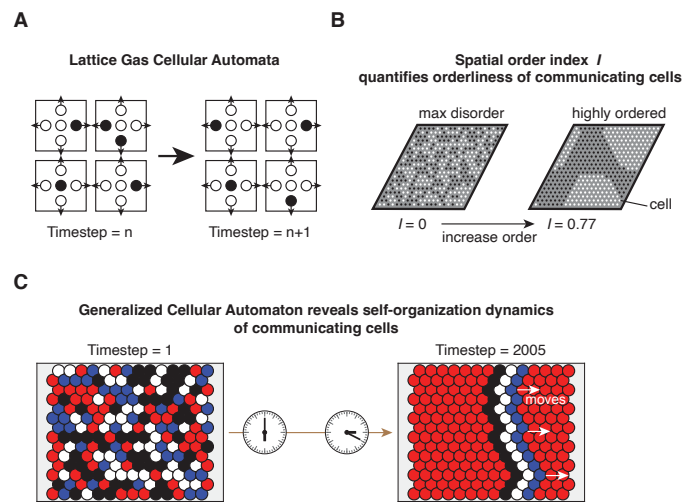

**Figure 4. Generalized cellular automata as a framework for multicellular self-organization.**
*(A) Lattice-gas cellular automaton (LGCA) setup on a square lattice, where each site contains discrete channels corresponding to possible velocity directions. Particles occupy channels and update synchronously through propagation and collision rules. (B) Quantification of spatial order using the spatial order index (I). Random arrangements yield I ≈ 0, while clustering of states produces higher I values, capturing emergent order in a single measure. (C) Generalized cellular automaton in which cells communicate by secreting and sensing two diffusible molecules. Local interactions yield dynamic spatial organization, here illustrated by a propagating linear wave pattern.*

In one class of generalized CA, extracellular molecules are modeled with reaction–diffusion equations[49,50]. Because gene-expression changes are typically much slower than molecular diffusion, one can assume the molecules equilibrate at each time step, then update cell states accordingly— this makes time, space, and states all discrete. This reliance on discrete updates parallels a challenge familiar from quantum many-body dynamics: once states are discrete, there is no single smooth density function that describes the system, and the number of possible configurations grows exponentially. Discreteness makes forward evolution tractable step by step but renders global prediction and backward inference overwhelmingly complex.

Within this framework, one study showed that cells secreting and sensing a single molecule under positive feedback can self-organize from disorder into a static, ordered pattern[49]. Each cell had two states—on (high secretion) or off (low secretion)—and order was quantified by a communication-weighted spatial order index I, a weighted covariance of cell states—a number between zero and one—that combines state similarity with communication strength (e.g., effective pairwise influence set by diffusion or uptake)[49,51]. Random arrangements yield I ~ 0, while clustering drives I upward, condensing the entire lattice configuration into a single number (Fig. 4B). Extending the rules to two secreted molecules—the I still applies here—produced a qualitatively richer repertoire: instead of static patterns, cells generated traveling waves, spirals, and other persistent dynamics, showing how additional communication channels enable not just order but its continual renewal[50]. These models illustrate how generalized CA can capture emergent behaviors inaccessible to nearest-neighbor automata (Fig. 4C). Moreover, these results show that rule-based systems, like living systems, can shift from sustained, renewing dynamics (e.g., waves) to static states once their underlying communication rules can no longer maintain coherent updates[50]—an emergent analogue of viability loss.

Other recent work has also employed CA to probe biological self-organization[52]. One study used one-dimensional automata in which discrete subunits updated their states according to local signaling rules, showing that programmable pattern formation can emerge from minimal ingredients without global control[52]. Another revisited the classical "French flag" problem, evolving one-dimensional CA rules to produce robust and tunable spatial boundaries without relying on global signals[53]. These studies demonstrate how even simple, linear arrangements of locally coupled rules can generate a surprising diversity of patterns. By contrast, generalized CA in two dimensions have revealed not only static structures but also dynamic spatial patterns such as spirals and traveling waves. Applications of CA to biological systems at this mechanistic level remain uncommon, with most CA research confined to abstract mathematics or computer science. This rarity highlights an underexplored opportunity: discrete, rule-based models provide a complementary grammar to the force- and energy-based frameworks of physics, and may be essential for capturing the kinds of self-organization dynamics unique to living systems.

## Outlook

The question of resilience—the ability of living systems to withstand perturbations without losing viability—versus viability itself raises a deeper issue: is the inevitability of death in all known organisms simply a consequence of how life has evolved, or does it reflect a more fundamental physical law that forbids any living system from sustaining its NEC indefinitely? Each organism—whether a minimal cell, yeast, or animal—embodies a unique genome and molecular architecture that define its specific resilience and fragility. In this sense, the NEC is inseparable from an organism's design: resilience and viability are two facets of the same capacity. Yet the broader question remains whether it is even possible to construct, through nature or design, a living system—perhaps one not based on DNA, that never loses its NEC. Comparative studies across diverse forms of life may eventually reveal the universal principles that both enable and constrain this finite capacity.

Living systems demand a rethinking of how physics approaches self-organization. Frameworks developed for non-living matter—equilibrium phase transitions, dissipative structures, even stochastic thermodynamics—have given profound insights, but they cannot by themselves explain why cells remain alive for finite spans of time and why their capacity for non-equilibrium dynamics is inevitably lost. Nor do they naturally accommodate the discreteness, rule-driven interactions, and computational logic that underlie processes like gene expression, cell signaling, and collective decision-making. Collectively, these clarifications strengthen the conceptual continuity between the experimental and computational approaches proposed here, linking the physical limits of NEC observed in cells with the rule-based limits explored through cellular automata.

The perspective presented here reflects one view of how to move forward, while acknowledging that there are certainly other promising directions and viewpoints. This Perspective emphasizes two complementary approaches. On one hand, slowing and suspending life processes offers an experimental foothold to probe the limits of viability and to clarify why death is both irreversible and inevitable. On the other, generalized cellular automata offer a computational framework to explore how discrete, local rules coalesce into global organization, revealing principles inaccessible to continuous, force-based descriptions. Together, these approaches highlight that living systems are not only far-from-equilibrium physical systems but also rule-executing, information-processing systems whose dynamics cannot be reduced to energies and fluxes alone.

The challenge ahead is to articulate a new grammar—one that integrates physics with computation, discreteness, and biological rules—to explain how self-organization both sustains life and, ultimately, fails. Such a framework would not replace the powerful tools of equilibrium or non-equilibrium physics but extend them, complementing force- and energy-based perspectives with rule- and computation-based ones. Just as concepts like entropy and temperature emerged to bridge micro- and macroscopic physics, new quantities may yet emerge to bridge molecules, cells, and organisms. Meeting this challenge would not only deepen our understanding of life's dynamics but also expand the reach of physics itself.

## ACKNOWLEDGEMENTS

HY was supported by a grant from the National Institutes of Health (NIH-NIGMS R35 Grant, GM147508).

## DECLARATION OF INTERESTS

The author declares no competing interests.